\documentclass[a4paper,11pt]{article}
\usepackage[utf8]{inputenc}

\usepackage[table]{xcolor}
\usepackage{jcappub}
\usepackage{graphicx}
\usepackage{psfrag,fancyhdr,epsfig}
\usepackage{hyperref}

\usepackage{cancel}
\usepackage{soul}

\usepackage{amsmath}
\usepackage{array}
\usepackage{subfig}
\usepackage{graphicx}
\usepackage{color}
\usepackage{tensor}
\usepackage{xcolor}
\usepackage{orcidlink}
\usepackage{comment}
\usepackage{physics}
\usepackage[mathscr]{eucal}
\usepackage{multirow}
\usepackage{graphicx}
\usepackage{amssymb}
\usepackage{amsmath}
\usepackage{subfig}
\usepackage{xcolor}

\newcommand{\be}{\begin{equation}}
\newcommand{\ee}{\end{equation}}
\newcommand{\bear}{\begin{array}}
\newcommand{\eear}{\end{array}}
\newcommand{\ba}{\begin{eqnarray}}
\newcommand{\ea}{\end{eqnarray}}

\usepackage{array}
\newcolumntype{C}[1]{>{\centering\arraybackslash}p{#1}}

\usepackage[most]{tcolorbox}

\title{\huge \centering Palatini inflationary (non-)attractors:\\ Universal mapping of observables}

\author[a]{Christian Dioguardi,}
\author[a,b]{Francesco Gianesello,}
 \author[a]{Antonio Racioppi}

\emailAdd{christian.dioguardi@kbfi.ee}
\emailAdd{antonio.racioppi@kbfi.ee} 
\emailAdd{francesco.gianesello@taltech.ee}

\affiliation[a]{Laboratory of High Energy and Computational Physics, 
National Institute of Chemical Physics and Biophysics, R{\"a}vala pst.~10, Tallinn, 10143, Estonia}

\affiliation[b]{Tallinn University of Technology, Akadeemia tee 23, 12618 Tallinn, Estonia}

\abstract{We study single-field slow-roll inflation in the context of Palatini gravity for the class of non-minimally coupled $\xi$-attractors, i.e.,\ models where the same function $f(\phi)$ fixes both the non-minimal coupling, $1+\xi f(\phi)$, and the inflationary potential, $V(\phi) = M^4 f(\phi)^2$. For an arbitrary $f(\phi)$, the number of $e$-folds is, at leading order, independent of $\xi$. Thanks to this, we provide an immediate mapping between the observables of the non-minimally coupled setup and those of the minimally coupled one. In the strong coupling limit, the tensor-to-scalar ratio is, as usual, suppressed, while the scalar spectral index is shifted towards larger values, with the magnitude of the shift depending only on the tensor-to-scalar ratio associated with the original potential $V(\phi)$.}

\begin{document}

\maketitle

\section{Introduction}

Cosmic inflation is widely considered the leading paradigm for the description of the primordial Universe: besides addressing the homogeneity and flatness problems, it generates the perturbations that seed the observed large-scale structure \cite{Starobinsky:1980te,Guth:1980zm,Linde:1981mu,Albrecht:1982wi,Mukhanov:1981xt}. The simplest realization relies on a single scalar field slowly rolling along a sufficiently flat potential. Most of its phenomenology can be encoded into a small number of CMB observables: the amplitude of the scalar power spectrum $A_s$, the spectral index $n_s$ and the tensor-to-scalar ratio $r$.

Since the combination provided by the BICEP/Keck collaboration~\cite{BICEP:2021xfz}, the allowed parameters space has been sensibly reduced, indicating as preferred realizations the Starobinsky model~\cite{Starobinsky:1980te} and its variations, like the $\xi$-attractors~\cite{Kallosh:2013tua, Kallosh:2013yoa,Galante:2014ifa}. Both scenarios can be described by a scalar field non-minimally coupled to gravity  (e.g.~\cite{Galante:2014ifa,Jarv:2016sow} and references therein). 

However, when theories exhibit non-minimal couplings to gravity, the choice of
the dynamical degrees of freedom becomes crucial. 
In the usual metric formulation the connection is fixed to be the Levi-Civita one, whereas in the Palatini formulation the metric and the connection are treated as independent variables \cite{Einstein:1925,Ferraris:1982,Sotiriou:2008rp}. The two formulations are equivalent for the Einstein-Hilbert action with minimally coupled matter, but the equivalence breaks as soon as the scalar field couples non-minimally to the Ricci scalar \cite{Koivisto:2005yc,Bauer:2008zj}, or when the gravitational action is extended beyond the Einstein-Hilbert term \cite{Enckell:2018hmo,Antoniadis:2018ywb}. The Palatini framework has been extensively studied in the context of inflation (e.g. \cite{Tamanini:2010uq,Bauer:2010jg,Rasanen:2017ivk,Tenkanen:2017jih,Racioppi:2017spw,Markkanen:2017tun,Jarv:2017azx,Racioppi:2018zoy,Kannike:2018zwn,Enckell:2018kkc,Rasanen:2018ihz,Bostan:2019uvv,Bostan:2019wsd,Carrilho:2018ffi,Almeida:2018oid,Takahashi:2018brt,Tenkanen:2019jiq,Tenkanen:2019xzn,Tenkanen:2019wsd,Kozak:2018vlp,Gialamas:2019nly,Racioppi:2019jsp,Rubio:2019ypq,Edery:2019txq,Lloyd-Stubbs:2020pvx,Das:2020kff,McDonald:2020lpz,Shaposhnikov:2020fdv,Enckell:2020lvn,Gialamas:2020snr,Karam:2020rpa,Gialamas:2020vto,Karam:2021wzz,Dioguardi:2021fmr,Karam:2021sno,Gialamas:2021enw,Annala:2021zdt,Racioppi:2021ynx,Kodama:2021yrm,Cheong:2021kyc,Mikura:2021clt,Ito:2021ssc,Racioppi:2021jai,Gialamas:2022gxv,Dioguardi:2022oqu,Dimopoulos:2022rdp,Dimopoulos:2022tvn,Gialamas:2022xtt,Bostan:2023ped,TerenteDiaz:2023kgc,Kannike:2023kzt,Dioguardi:2023jwa,Dimopoulos:2025fuq,Gialamas:2024uar,Racioppi:2024pno,Bostan:2024ugi,Dioguardi:2025mpp,Dioguardi:2025vci,Bostan:2025dej,Bostan:2025xrx,Pallis:2025nrv,Kallosh:2026tjm,Kallosh:2026qrc} and refs therein).
It is well-known that in this framework a non-minimal coupling to the Ricci scalar can suppress the tensor-to-scalar ratio by orders of magnitude. Moreover, a Palatini version of $\xi$-attractors, i.e., models in which the same function $f(\phi)$ fixes the non-minimal coupling, $1+\xi f(\phi)$, and the inflationary potential, $V(\phi) = M^4 f(\phi)^2$, was also discussed in \cite{Jarv:2017azx,Kodama:2021yrm,Pallis:2025nrv}.

In the metric formulation this construction is known to reproduce the Starobinsky predictions irrespective of the choice of $f(\phi)$ \cite{Kallosh:2013tua,Kallosh:2013yoa,Galante:2014ifa}. The Palatini counterpart behaves in a substantially different way. The number of $e$-folds is independent of $\xi$ at the leading order, therefore the CMB observables can be approximated in terms of their minimally coupled counterparts, with $r$ and $A_s$ suppressed by the same factor and $n_s$ (model-dependently) increased. The paper is organized as follows. In section \ref{sec:model} we introduce the setup and define the Jordan and Einstein frame actions. In section \ref{sec:observables} we derive the inflationary observables for a generic $f(\phi)$ in the slow-roll approximation and test our approach against the previous results of \cite{Jarv:2017azx}. In section \ref{sec:conclusions} we present our conclusions.

\section{Non-minimally coupled Palatini action}\label{sec:model}
We start from a Jordan frame action for a real scalar field $\phi$ non-minimally coupled to  Palatini gravity as:
\be\label{eq:action}
S_J = \int d^4x \sqrt{-g_J} \qty[\frac{M_P^2}{2}\Big(1 + \xi f(\phi)\Big) R_J -\frac{1}{2}g^{\mu\nu}_J\partial_\mu\phi\partial_\nu\phi - V(\phi)]\,,
\ee
where $M_P$ is the reduced Planck mass, $ f(\phi)$ is the generalized (dimensionless) non-minimal coupling function, $\xi$ a dimensionless coupling,  $g^{\mu\nu}_J$ the Jordan frame metric and the scalar field potential is given by
\be\label{eq:potential}
V(\phi) \equiv M^4 f(\phi)^2\, ,
\ee
where $M$ has the dimensions of a mass.
Notice that, since we are in the Palatini formalism, $R_J = g^{\mu\nu}_J R_{\mu\nu}(\Gamma)$ is constructed from the Palatini Ricci tensor which is by construction independent of the metric.
Then, after a conformal redefinition of the metric,
\be\label{eq:weyl}
g_{\mu\nu}^E = \Omega^2(\phi) g_{\mu\nu}^J \equiv \Big(1+\xi f(\phi)\Big) g_{\mu\nu}^J\,,
\ee
the action takes the equivalent form\footnote{To ensure that the Weyl transformation \eqref{eq:weyl} is well-defined everywhere in field space, it is necessary to impose $1+\xi f(\phi)>0$ for every $\phi$. From now on, we will assume $\xi>0$ and $f(\phi)>0$.} in the Einstein frame:
\be
S_E = \int d^4x \sqrt{-g_E}\qty[\frac{M_P^2}{2}R_E - \frac{1}{2\Omega(\phi)^2}g^{\mu\nu}_E\partial_\mu\phi\partial_\nu\phi - U(\phi)]\,,
\ee
with
\be
U(\phi) \equiv \frac{M^4 f(\phi)^2}{\left(1+\xi f(\phi)\right)^2}\,.
\ee
By introducing a field redefinition
\be\label{eq:canonical_scalar}
\qty(\frac{\partial\chi}{\partial\phi})^2 = \frac{1}{1+\xi f(\phi)}\,,
\ee
we can rewrite the action in terms of the canonical Einstein frame scalar field
\be\label{eq:action_einstein}
S_E = \int d^4x \sqrt{-g_E}\qty[\frac{M_P^2}{2}R_E - \frac{1}{2}g^{\mu\nu}_E\partial_\mu\chi\partial_\nu\chi - U(\chi)]\,.
\ee
\section{Slow-roll inflation and observables}\label{sec:observables}
Starting from \eqref{eq:action_einstein} we can use the slow-roll formalism to compute the main CMB observables. We first define the slow-roll parameters:
\begin{align}\label{eq:slow-roll}
 \epsilon(\chi) &\equiv \frac{M_P^2}{2}\left(\frac{ U'(\chi)}{U(\chi)} \right)^2\,, \\  
 \eta(\chi) &\equiv M_P^2\left(\frac{U''(\chi)}{U( \chi)} \right)\,,
\end{align}
where $'$ represents a derivative with respect to the argument of the function. 
In the slow-roll approximation, these can be related to the main CMB observables for inflation, i.e., the tensor-to-scalar ratio $r$, the spectral index $n_s$, the amplitude of the scalar perturbation $A_s$ and the number
of $e$-folds at horizon crossing $N_*$:
\begin{align}\label{eq:CMB_obs}
r &= 16\epsilon(\chi_*)\,,\\
n_s &=1-6\epsilon(\chi_*)+2\eta(\chi_*)\,, \\
A_s &= \frac{1}{24\pi^2}\frac{U(\chi_*)}{M_P^4}\frac{1}{\epsilon(\chi_*)}\,,\\
N_* &=\frac{1}{M^2_P}\int_{\chi_{\rm{end}}}^{\chi_*}d\chi \frac{U(\chi)}{U'(\chi)}\, ,   \label{eq:Ne:general}
\end{align}
where $\chi_\text{end}$ is defined by $\epsilon(\chi_\text{end})=1$. 
For a general $f(\phi)$ it is not possible to integrate analytically \eqref{eq:canonical_scalar} and obtain explicit expressions in terms of $\chi$. This is however not necessary to compute the slow-roll parameters and CMB observables in terms of $f(\phi)$ and its derivatives.
Indeed, using the chain rule for derivatives one can compute that:
\begin{align}
\epsilon(\phi) &=\frac{2 M_P^2f'(\phi)^2}{f(\phi)^2(1+\xi f(\phi))}\,, \label{eq:eps}\\
 \eta(\phi)
&= \frac{M_P^2(2-3\xi f(\phi))f'(\phi)^2}{(1+\xi f(\phi))f(\phi)^2} +\frac{2 M_P^2 f''(\phi)}{f(\phi)}\,,
\end{align} 
which give
\begin{align}
r &=\frac{32 M_P^2 f'(\phi_*)^2}{f(\phi_*)^2(1+\xi f(\phi_*))}\,,\\
n_s&=1+2M_P^2\left[-\frac{(4+3\xi f(\phi_*))f'(\phi_*)^2}{(1+\xi f(\phi_*))f(\phi_*)^2} +\frac{2f''(\phi_*)}{f(\phi_*)} \right]\,, \\
A_s &= \frac{M^4 f(\phi_*)^4}{48\pi^2M_P^6 (1+\xi f(\phi_*))f'(\phi_*)^2}\,,\\ \label{eq:e-folds_phi}
N_* &= 
\frac{1}{2 M_P^2}\int_{\phi_{\rm{end}}}^{\phi_*}d\phi \frac{f(\phi)}{f'(\phi)}\,. 
\end{align} 
The integral \eqref{eq:e-folds_phi} cannot be evaluated until the explicit form of $f(\phi)$ is known, however we can see that it is independent on $\xi$, except for the exact value of $\phi_\text{end}$ defined by $\epsilon(\phi_\text{end})=1$ (see eq. \eqref{eq:eps}). On the other hand, it is well-known that  the contribution coming from $\phi_\text{end}$ is subdominant in the numerical evaluation of $N_*$. Therefore we can conclude that, at the leading order (i.e. ignoring the contribution from $\phi_\text{end}$), the evaluation of $N_*$ (and therefore $\phi_*$) remains the same for any value of $\xi$, including $\xi=0$. Rephrasing it, at the leading order, the formal expression of $N_*$ at any $\xi$ coincides with the one in the case $\xi = 0$. This last point can be proven as follows:
 \begin{eqnarray}
 \bar N_* 
 &=&  \frac{1}{M_P^2}\int_{\bar\phi_\text{end}}^{\bar\phi_*}d\phi \frac{V(\phi)}{V'(\phi)} 
 =  \frac{1}{M_P^2}\int_{\bar\phi_\text{end}}^{\bar\phi_*}d\phi f(\phi)^2 \left[\frac{\partial \left( f(\phi)^2 \right)}{\partial \phi}\right]^{-1}
  \nonumber\\
 &=& \frac{1}{M_P^2}\int_{\bar\phi_\text{end}}^{\bar\phi_*}d\phi f(\phi)^2 \left[ 2 f(\phi) f'(\phi)\right]^{-1} 
 =\frac{1}{2 M_P^2}\int_{\bar\phi_\text{end}}^{\bar\phi_*}d\phi \frac{f(\phi)}{f'(\phi)} \, , \label{eq:N0}
 \end{eqnarray}
where a bar indicates a quantity evaluated at $\xi=0$. This is a well-known result (e.g. \cite{Bauer:2008zj,Carrilho:2018ffi,Kodama:2021yrm} and refs. therein). The expression in eqs. \eqref{eq:e-folds_phi} and \eqref{eq:N0} are formally equivalent, however a subtle dependency on $\xi$ still remains via the contribution of the field value at the end of inflation, determined by solving $\epsilon=1$, which is explicitly depending on $\xi$. However, it is also well-known that such a contribution is subdominant, therefore we can neglect it at the leading order and approximate \eqref{eq:e-folds_phi} as $N_* \simeq \bar N_*$ and therefore $\phi_* \simeq \bar\phi_*$.
This implies that we can express the CMB observables in terms of the observables in absence of non-minimal coupling, $\bar r, \bar n_s, \bar A_s$. We find\footnote{
To the best of our knowledge, it is the first time in the literature that such a mapping between the observables is shown in detail. In particular, in \cite{Kodama:2021yrm}, the equations for the observables are expressed in terms of the slow-roll parameters of the original potential but neither an explicit approximation $\phi_* \simeq \bar\phi_*$  (at all $\xi$ values) nor a direct connection between observables is presented. Additionally, the universal positive shift of the scalar spectral index and its upper bound are not identified there.
} that
\begin{align}
  \label{eq:rr0}  r &\simeq \frac{\bar r}{1+\xi f(\bar\phi_*)}\,,\\ \label{eq:nsn0}
    n_s &\simeq \bar n_s +  \frac{2 \xi M_P^2  f'(\bar\phi_*)^2}{f(\bar\phi_*) \big( 1+\xi f(\bar\phi_*) \big)} = \bar{n}_s + \frac{\bar r-r}{16} \,, \\
    A_s &\simeq\frac{\bar A_s}{1+\xi f(\bar\phi_*)}= \bar A_s \frac{r}{\bar r} \,, 
\end{align}
and that for $\xi \rightarrow \infty$ the model generally predicts:
\begin{align}
r &\simeq 0\,, \label{eq:rasympt} \\
n_s &\simeq \bar n_s + 2M_P^2 \left( \frac{f'(\bar\phi_*)}{f(\bar\phi_*)}\right)^2 = \bar{n}_s + \frac{\bar r}{16} \, .  \label{eq:nsasympt}
\end{align}
Eqs. \eqref{eq:rr0} and \eqref{eq:rasympt} imply that $r$ is a decreasing function of $\xi$ approaching asymptotically zero. On the other hand, eqs. \eqref{eq:nsn0} and \eqref{eq:nsasympt} imply that $n_s$ is an increasing function of $\xi$ approaching asymptotically the value given in eq. \eqref{eq:nsasympt}. We stress that such a value depends only on $\bar r$ and $\bar n_s$, which are the observables of the original potential $V(\phi)$.

It is useful to compare our results with those of the original paper \cite{Jarv:2017azx}, where $f$ was a monomial function: $f(\phi)=\left(\frac{\phi}{M_P}\right)^{n}$.

The result of eq. \eqref{eq:rasympt}, is trivially checked. Therefore we need only to reproduce the result concerning $n_s$.
Using equation \eqref{eq:e-folds_phi} and neglecting the subdominant contribution of the field at the end of inflation, it is easily proven that
\begin{eqnarray}
    \bar r &\sim&\frac{8n}{N_*} \, , \\
    \bar n_s &\sim&1-\frac{n+1}{N_*} \, .
\end{eqnarray}
Inserting the equations above into \eqref{eq:nsasympt}, we obtain
\begin{equation}
    n_s\sim 1-\frac{n+1}{N_*}+\frac{8n}{16 N_*}=1 -\left( 1+\frac{n}{2}\right)\frac{1}{N_*}\,.
\end{equation}
This result coincides with the one obtained with a completely independent approach in \cite{Jarv:2017azx}, further validating our method and approximation.

\section{Conclusions}\label{sec:conclusions}

We have considered slow-roll inflation in the framework of Palatini gravity for a single real scalar field non-minimally coupled to the Ricci scalar, restricting ourselves to the class of $\xi$-attractors, where the same function $f(\phi)$ defines both the non-minimal coupling $1+\xi f(\phi)$ and the potential $V = M^4 f(\phi)^2$. The starting point is that in the Palatini formulation the number of $e$-folds \eqref{eq:e-folds_phi} does not depend on $\xi$ and coincides with the one of the minimally coupled model (neglecting the contribution at the end of inflation). This allows us to express, for an arbitrary $f(\phi)$, the CMB observables in terms of their $\xi=0$ counterparts: the tensor-to-scalar ratio and the amplitude of the scalar perturbations are suppressed by the same factor, $r = \bar r/(1+\xi f(\bar\phi_*))$ and $A_s = \bar A_s/(1+\xi f(\bar\phi_*))$, while the spectral index is always shifted towards larger values, $n_s > \bar n_s$, with an upper bound corresponding to the $\xi \to \infty$ limit given in eq. \eqref{eq:nsasympt}. Such a limit depends only on the values $\bar n_s$ and $\bar r$ predicted by the original potential $V(\phi)$.
Therefore, by increasing $\xi$ we can arbitrarily decrease $r$ but not $n_s$, which asymptotically approaches the value \eqref{eq:nsasympt}. According to our knowledge, the explicit mapping between observables is a new result never shown in the existing literature.
 
Unlike in the metric formulation, where the strong coupling limit of $\xi$-attractors is universal and reproduces the Starobinsky predictions \cite{Kallosh:2013tua}, here there is an explicit dependence on the original potential $V$, which directly translates into the model-dependent attractor value of $n_s$. However, such a mechanism can be very attractive for inflationary model building, regardless of the \emph{winner} of the CMB-BAO \emph{tension} \cite{AtacamaCosmologyTelescope:2025nti,Ferreira:2025lrd,Balkenhol:2025wms}: models, that in the minimally coupled case would predict a too small $n_s$ and too large $r$, might be brought into agreement with the data by increasing the $\xi$ parameter alone. Moreover, the predictions for the strong coupling limit can be easily derived from the predictions of the original model. 

We conclude that non-minimally coupled Palatini $\xi$-attractors provide a simple and general mechanism for viable inflation, and predict a model-dependent value of $n_s$ which can be discriminated by next-decade CMB observations such as the Simons Observatory, LiteBIRD and SPT 3G+~\cite{SimonsObservatory:2018koc,LiteBIRD:2022cnt,SPT:2026olz}.

\acknowledgments

This work was supported by the Estonian Research Council grants PRG1677, TARISTU24-TK10, TARISTU24-TK3, and by the CoE program TK202 ``Foundations of the Universe''. This article is based upon work from COST Action CosmoVerse CA21136, supported by COST (European Cooperation in Science and Technology).

\bibliography{references}

\end{document}